\documentstyle[aasms4]{article}
\begin{document}
\title{Probing the two temperature paradigm for advection dominated
accretion flow: test for the component thermalization
time-scale passed.}
\author{David Tsiklauri}
\affil{Physics Department, Tbilisi State University,
3 Chavchavadze Ave., 380028 Tbilisi, Georgia.
email: dtsiklau@resonan.ge}
\begin{abstract}
We report here on a calculation of thermalization time-scale
of the two temperature advection dominated accretion flow (ADAF)
model. It is established that time required to equalize the electron
and ion temperatures via electron-ion collisions
in the ADAF with plausible physical parameters
greatly exceeds age of the Universe, which corroborates validity
one of the crucial assumptions of the ADAF model, namely
the existence of a hot {\it two temperature plasma}.
This work is motivated by the recent success (Mahadevan 1998a,b) of
ADAF model in explaining the emitted spectrum of Sgr A$^*$.
\end{abstract}

Identification of the nature of enigmatic radio source
Sgr A$^*$ at the Galactic center
has been source of debates since its discovery. Observations of
stellar motions at the Galactic center
(Eckart \& Genzel, 1997; Genzel et al., 1996)
and low proper motion ($\leq$ 20 km sec$^{-1}$; Backer, 1996)
of  Sgr A$^*$ indicate that, on the one hand,
it is a massive $(2.5 \pm 0.4) \times 10^6 M_\odot$
object dominating the gravitational potential in the
inner $\leq 0.5$ pc region of the galaxy. On the other hand,
observations of stellar winds and other gas flows in the vicinity
of Sgr A$^*$ suggest that the mass accretion rate $\dot M$ is about
$6 \times 10^{-6}M_\odot$yr$^{-1}$ (Genzel et al., 1994).
This implies that the luminosity
of the central object should be more than $10^{40}$ erg sec$^{-1}$,
provided the radiative efficiency is the usual 10\%.
However, observations indicate that the
bolometric luminosity is actually less than $10^{37}$ erg sec$^{-1}$.
This discrepancy has been a source of exhaustive debate in the
recent past.

The broad-band emission spectrum of  Sgr A$^*$
can be reproduced either in the quasi-spherical accretion model
(Melia, 1992, 1994) with $\dot M \simeq 2 \times 10^{-4}M_\odot$
yr$^{-1}$ or by a combination of disk plus radio-jet model
(Falcke et al., 1993a, 1993b).
As pointed out by Falcke and Melia (1997), quasi-spherical accretion
seems  unavoidable at large radii, but the low actual luminosity
of Sgr A$^*$ points toward a much lower accretion rate in a starving
disk. Therefore, Sgr A$^*$ can be described by a model of a
fossil disk fed by quasi-spherical accretion.
Recently, Tsiklauri \& Viollier  (1998) have proposed
an alternative model for the mass distribution at the galactic
center in which the customary supermassive
black hole is replaced by a ball composed of self-gravitating,
degenerate neutrinos. It has been shown that
a neutrino ball with a mass $2.5 \times 10^6 M_\odot$,
composed of neutrinos and antineutrinos with  masses
$m_\nu \geq 12.0$ keV$/c^2$
for $g=2$ or $m_\nu \geq 14.3$ keV$/c^2$ for $g=1$, where $g$
is the spin degeneracy factor, is  consistent with the
current observational data. See also Munyaneza, Tsiklauri and
Viollier 1998 for future tests of the model.
Tsiklauri \& Viollier 1999 have performed calculations of the spectrum
emitted by Sgr A$^*$ in the framework of standard accretion disk theory,
assuming that Sgr A$^*$ is a neutrino ball with the above mentioned
physical properties, and established that at least part of the
calculated spectrum, were the observational data is most reliable,
is consistent with the observations.

Probably the most successful model which is consistent with the
observed emission spectrum of Sgr A$^*$ has been developed
by Narayan et al., 1995, 1998 (see also Manmoto et al., 1997).
This model is based on the concept of advection dominated accretion
flow (ADAF), in which most of the energy released by viscosity in
the disk is carried along with the plasma and lost into the black hole,
while only a small fraction is actually radiated off.
Recent papers by Mahadevan 1998a,b have significantly advanced
ADAF model. Inclusion of additional emission component, namely
synchrotron radiation from $e^{\pm}$ created via decay of
charged pions, which in turn are produced through proton-proton
collisions in the ADAF, has significantly improved fitting
of the Sgr A$^*$ spectrum in the low frequency band.
After removing the latter discrepancy ADAF model of the
Sgr A$^*$, apart from the size versus frequency constraints
(Lo et al. 1998 and references therein)
which remain problematic for all current emission models of
the radio source anyway, seems to be the most viable alternative.
Thus, basic assumptions of the ADAF model should be carefully
examined from the point of view physical consistency.
As appropriately pointed out by Mahadevan (1998b),
in order for the ADAF solutions to exist two basic
assumptions in plasma physics must be satisfied: (a) existence
of a hot {\it two temperature} plasma, and (b) the viscous
energy generated primarily heats the protons. As to the
assumption (b) its validity would be hard to verify as it
is related to the yet unknown mechanism of viscosity in the
accretion flow. The latter problem stands on its own in
astrophysics. As concerns the assumption (a) it is from
the field of plasma physics, a branch of physics which
has been expensively studied in the laboratory, were we
seem to have more comprehensive in comparison to astrophysics
understanding of the underlying
basic physical phenomena. Thus, motivated by the recent success
(Mahadevan 1998a,b) of ADAF model in explaining the emitted
spectrum of Sgr A$^*$, we set out with the aim to
check validity of the assumption (a).

It would be reasonable to believe that if the time
required to equalize the electron and ion temperatures
appears to be sufficiently large, then one might be
confident that the assumption of the existence of the
two temperature plasma in the ADAF is physically justified.
The relevant time scale for the temperature
equalization can be calculated using well formulated
methods know in plasma physics. The rate at which
temperature equilibrium between the electrons and ions
is approached is determined by (see e.g. Melrose, 1986):
$$
{{dT_e}\over{dt}}=\nu^{(e,i)}_{eq}(T_i-T_e), \eqno(1)
$$
$$
{{dT_i}\over{dt}}=-\nu^{(i,e)}_{eq}(T_i-T_e), \eqno(2)
$$
with
$$
\nu^{(e,i)}_{eq}={{e^2q^2_i n \ln \Lambda^{(e,i)}}
\over{3 {(2 \pi)}^{1/2} \pi m_e m_i \varepsilon^2_0}
(V_e+V_i)^{3/2}}.  \eqno(3)
$$
Here $\ln \Lambda^{(e,i)}$ is the Coulomb logarithm for
electron-ion collisions given by
$\ln \Lambda^{(e,i)}=22.0 - 0.5 \ln n_e + \ln T_e$,
($T_e > 1.4 \times 10^5$ K), $e$ and $q_i$ are charges of
electrons and ions respectively, $m_e$, $m_i$  and
$T_e$, $T_i$ are their masses and temperatures (in Kelvin),
$V_e$ and $V_i$ are thermal velocities of the electrons and ions,
$n$ is the number density of plasma
(we have assumed the global charge
neutrality of the ADAF, i.e. $n=n_e=n_i$)
and finally as the SI system of units is used,
$\varepsilon_0= 8.8541878 \times 10^{-12}$ F$/$m.

Now, writing the thermal velocities $V_{e,i}$ as
$V_{e,i}=\sqrt{T_{e,i}/m_{e,i}}$  in the Eqs.(1) -- (3),
these become closed set of ordinary differential equations
for $T_e$ and $T_i$. We set $q_i=e$ and $m_i$ as a mass of
proton. We solve numerically Eqs.(1) -- (3) using
{\tt Numerical Recipes Software}, namely {\tt odeint} driving
routine with fifth order
Cash-Karp Runge-Kutta method (tolerance error $10^{-15}$).
Calculations were performed for the values of the number densities
ranging from $10^{16}$ to $10^{31}$ m$^{-3}$.
The lower end of the number density's range is taken according to
the actual value of the $n$ in the ADAF
around Sgr A$^*$ (Manmoto 1997, 1999).
In the Fig.1 the number density (in cm$^{-3}$) profile is
plotted. As established by Manmoto (1997, 1999) such
ADAF number density profile corresponds to the case
when best fit of the ADAF model to the observed emission
spectrum is achieved. We gather form this plot
that the maximal value of the $n$ actually attained is somewhat
less than $10^{10}$ cm$^{-3}$ --- the value we use in our
calculations. Naturally, as more
dilute plasma is as more time will be required to
equilibrate electron and ion temperatures via
electron-ion collisions. Therefore, actual time
of the temperature equilibration is even larger than
those values obtained here.

The results of our calculations are presented in the
Fig. 2. For the initial values of the temperatures we
have used $T_e=10^{9.5}$ K and $T_i=10^{12}$ K respectively
(Mahadevan 1998a,b).
We gather from the plot that the time required to
equalize the temperatures of ions and electrons significantly
exceeds the age of the Universe. Therefore we conclude that
the assumption of the existence of a hot {\it two temperature} plasma
is valid, or at least initial temperature difference will not
be washed out by the electron-ion collisions within the age of the
Universe.

The only concern one has bear in mind is
that the formulas used in this paper, strictly speaking,
are valid for non-relativistic plasma regime (Melrose, 1999)
while the electron
and ion temperatures concerned are relativistic (for the electrons
$\gamma \simeq 200$). However, the temperature equalization
time-scale obtained is so large, that even inclusion of the
relativistic effects in our estimates would not
change basic besults of this paper drastically.

I would like to thank T. Manmoto (Kyoto University) for
calculating the ADAF number density profile and kindly
providing the Fig. 1. Also, I am thankful to
D. Melrose (University of Sydney) for providing the exact reference
of his book and useful comments on its contents, and to
R. Mahadevan (IoA, Cambridge)
for clarifying to me some points of the ADAF model.

\newpage
\centerline{figure captions:}
Fig. 1: The number density ($n$) profile which
corresponds to the case when best fit of the ADAF model
(Manmoto 1997, 1999) to the observed emission spectrum
of the Sgr A$^*$ is achieved.

Fig. 2: Solutions of the Eqs.(1) -- (3) for the three
values of the number density (log-log plot).
Thin lines correspond to
$n=n_e=n_i= 10^{31}$ m$^{-3}$, thick lines correspond to
$n=n_e=n_i= 10^{26}$ m$^{-3}$, while the thickest lines
correspond to $n=n_e=n_i= 10^{16}$ m$^{-3}$.
Solid lines correspond to the $T_e(t)$'s whereas
dashed lines correspond to the $T_i(t)$'s.
Note that equalization of the temperatures occurs
at times greatly exceeding age of the Universe ($\simeq 10^{10}$ yr)
and decreasing of the number density postpones the temperature
equalization which is, of course, in accordance to the general
physical expectations.
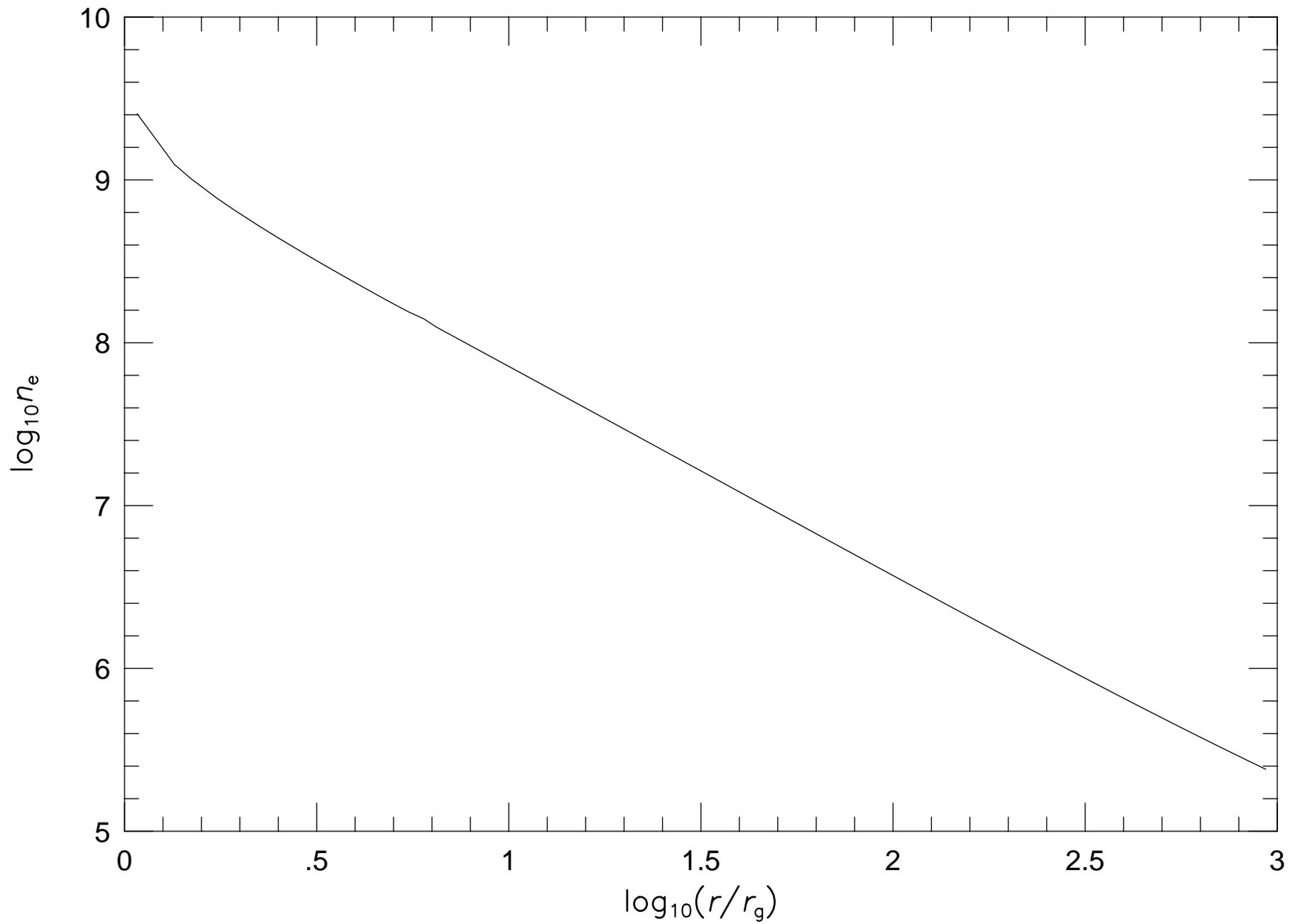
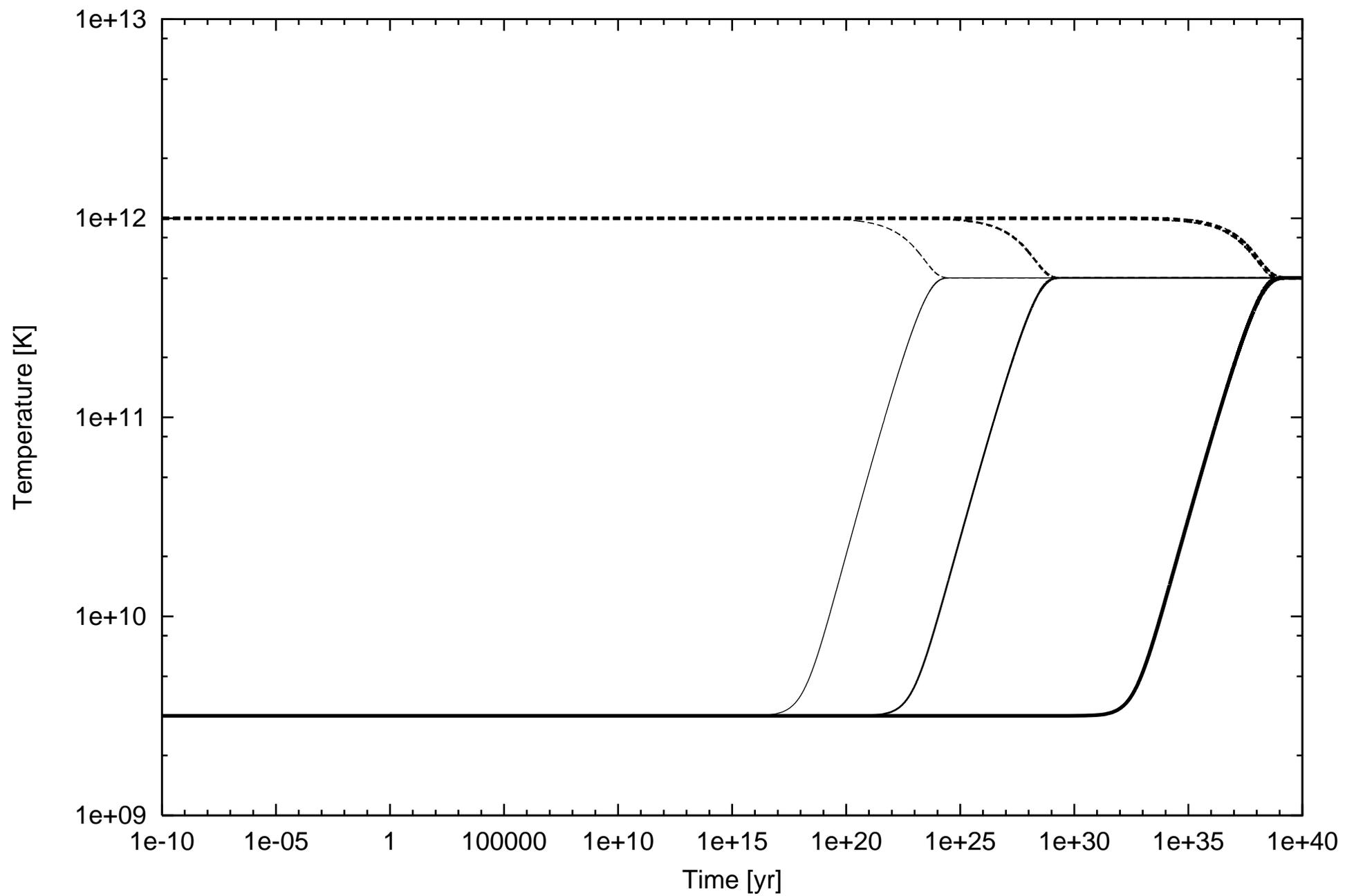
\end{document}